\documentclass[12pt]{article}
\usepackage{amsmath}
\usepackage{graphicx}
\usepackage{enumerate}
\usepackage{float}
\usepackage{amsfonts}
\usepackage{amsmath}
\usepackage{amssymb}
\usepackage{url} % not crucial - just used below for the URL 
\newtheorem{theorem}{Theorem}
\newenvironment{proof}[1][Proof]{\noindent \textbf{#1.} }{\  \rule{0.5em}{0.5em}}

\newcommand{\blind}{1}

\begin{document}

\def\spacingset#1{\renewcommand{\baselinestretch}%
{#1}\small\normalsize} \spacingset{1}

%%%%%%%%%%%%%%%%%%%%%%%%%%%%%%%%%%%%%%%%%%%%%%%%%%%%%%%%%%%%%%%%%%%%%%%%%%%%%%

\if1\blind
{
  \title{\LARGE\bf Estimating Viral Genetic Linkage Rates in the Presence of Missing Data}
  \author{Tyler Vu\thanks{
    The authors gratefully acknowledge \textit{NIAID R37 51164}}\hspace{.2cm}\\
    University of California, San Diego\
     \\
    Tuo Lin \\
    University of California, San Diego \ 
    \\
    Jingjing Zou \\ 
    University of California, San Diego \ 
    \\
    Vladimir Novitsky \footnotemark[1]\hspace{.2cm}\\
    Harvard University \ 
    \\ 
    Xin Tu \\ 
    University of California, San Diego \ 
    \\
    Victor De Gruttola \footnotemark[1]\hspace{.2cm}\\
    Harvard University}
  \maketitle
} \fi

\if0\blind
{
  \bigskip
  \bigskip
  \bigskip
  \begin{center}
    {\LARGE\bf Title}
\end{center}
  \medskip
} \fi

\bigskip
\begin{abstract}
Although the interest in the the use of social and information networks has grown, most inferences on networks assume the data collected represents the complete. However, when ignoring missing data, even when missing completely at random, this results in bias for estimators regarding inference network related parameters. In this paper, we focus on constructing estimators for the probability that a randomly selected node has node has at least one edge under the assumption that nodes are missing completely at random along with their corresponding edges. In addition, issues also arise in obtaining asymptotic properties for such estimators, because linkage indicators across nodes are correlated preventing the direct application of the Central Limit Theorem and Law of Large Numbers. Using a subsampling approach, we present an improved estimator for our parameter of interest that accommodates for missing data. Utilizing the theory U-statistics, we derive consistency and asymptotic normality of the proposed estimator. This approach decreases the bias in estimating our parameter of interest. We illustrate our approach using the HIV viral strains from a large cluster-randomized trial of a combination HIV prevention intervention - the Botswana Combination Prevention Project (BCPP).\end{abstract}
 
\noindent%
\vfill

\newpage
\spacingset{1.5} % DON'T change the spacing!
\section{Introduction}
\label{sec:intro}
\textbf{REMINDER FOR VICTOR: SOMEWHERE IN THIS SECTION YOU SAID YOU WOULD ADD SOMETHING ABOUT THIS WORK'S APPLICATIONS TO COVID}

While networks have become widely used to analyze elements in a system and how these elements interconnect, the challenge of sampling complete network data remains a prevalent issue. In most instances, we only sample a portion of the nodes and hence don't observe the edges corresponding to the missing nodes. As a result, estimators for linkage rates that ignore the impact of missing data will be biased downwards. \\
\indent Additionally, obtaining asymptotic properties for estimators of linkage rates is challenging, because linkage indicators across pairs of individuals are correlated; hence, the central limit theorem and law of large numbers cannot be directly applied to such estimators. \\
\indent For our parameter of interest, past work has been explored to accommodate missing data in the case of viral genentic linkage networks (which in turn would apply to networks general). In Liu et. al., a multiple imputation framework in which the missing sequences are imputed is used to adjust for the bias in estimation of linkage rates across individuals that results from the missing data \cite{Liu2015}. Carnegie et. al. consider a subsampling approach to develop such an adjustment \cite{Carnegie2014}. Neither of these papers demonstrate desirable asymptotic properties such as consistency and asymptotic normality. \\
\indent In this paper, the overall goal is to develop estimators for linkage rates under the assumption that unobserved nodes are missing completely at random (MCAR). First, we show that the bias can, under the MCAR assumption, be represented as a multiplicative factor equal to the probability that we observe a node's edge in the sample. From estimates of this factor, we construct an improved estimator for this multiplicative factor using a subsampling approach. A U-Statistics approach facilitates development of an improved estimator for linkage probabilities that are asymptotically normal. We refer to the proposed estimator as the adjusted estimator. Lastly, we propose a diagnostic approach for assessing the reliability of the method. \\
\indent We apply these methods to analyses of HIV viral genetic linkage network in Botswana where the data is from a large cluster-randomized trial of a combination HIV prevention intervention - the Botswana Combination Prevention Project (BCPP) \cite{novitsky2020}. The interest in these analyses is to investigate the patterns of HIV transmission between communities in Botswana.  \\
\indent The paper is organized as follows. Section 2 introduces the notation and setting along with our parameter of interest. Section 3 illustrates the bias of an estimator for $\theta_{rs}$ that arises when we do not adjust for incomplete data. Section 4 shows our proposed approach to adjust for incomplete data. Section 5 demonstrates the proposed approach applied to a simulation setting and the HIV viral genetic network from the BCPP. In Section 6, we discuss the overall findings from the proposed approach.

\section{Notation and Setting}
Consider a population of nodes, $\Omega$, of finite size $N$  partitioned into $w$ disjoint groups, $\Omega_{1},\ldots,\Omega_{w}$, with $N_{r}$ being the number of nodes in group $r$.  Let
$\mathbf{y}_{ri}$ denote the $i$th node in group $r$. Let a network to be represented by $G = (\Omega, E)$ where $E$ is the set of edges between nodes, $E \subset \Omega \times \Omega.$ Note that we assume that $G$ is an undirected network. \\
\indent Let $D_{ri}^s$ be the number of edges that $\mathbf{y}_{ri}$ has in $\Omega_s$ (excluding $\mathbf{y}_{ri}$ if $r = s$). Let $N_{rs} = N_r + N_s$ if $r \neq s$. Otherwise, let $N_{rs} = N_r.$ We assume that the nodes and edges in $\Omega$ come from some network generating process and that $N_{rs}$ is sufficiently large such that the network structure of $\Omega$ is that of its network generating process. Also, we assume that  $\frac{\text{max}(D_{r1}^s, \ldots, D_{rN_r}^{s})}{\sqrt{N_{rs}}} \rightarrow 0$ as $N_{rs} \rightarrow \infty$. \\
\indent We are interested in inference about the probability that for a randomly selected node in group $r$ there exists at least one edge to some node in group $s$ (excluding itself if $r = s$), and we  refer to it as the linkage rate. We denote the linkage rate as the following: 

\begin{align*}
    \theta_{rs} = \Pr(D_{ri}^s \geq 1)
\end{align*}

\noindent where $\Pr()$ is defined by the superpopulation of infinite size. In practice, we need to estimate $\mathbf{\theta}_{rs}$ based on a sample from
$\Omega$.  Consider a random sample of subjects  from the nodes in $\Omega$ of size $n$,
which we denote by $S_{n}$, such that the proportion of sampled subjects in group $r$ is
$p_{r}$ (known). We denote the sample from $\Omega_{r}$
as $S_{n(r)}$ and the size of $S_{n(r)}$ as $n_{r}$. Then $n=\sum_{r=1}%
^{w}p_{r}N_{r}$.

\section{Bias Arising from Incomplete Data}

\begin{figure}
  \begin{minipage}[b]{0.60\linewidth}
  \vspace{0pt}
    \centering
    \includegraphics[scale = .7]{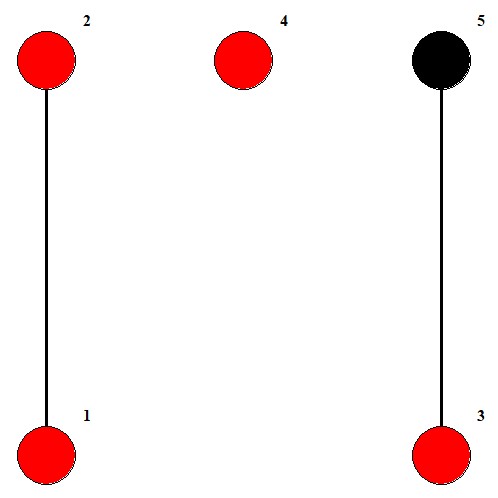}
  \end{minipage}%
  \begin{minipage}[b]{0.30\linewidth}
  \vspace{0pt}
    \centering
\begin{tabular}{|l|l|l|}
\hline
i  & $v_{ri}^r$ & $u_{ri}^r$ \\ \hline
1 & 1                            & 1                            \\ \hline
2 & 1                            & 1                            \\ \hline
3 & 0                            & 1                            \\ \hline
4 & 0                            & 0                            \\ \hline
\end{tabular}
\
\end{minipage}
\caption{Plot of a network to show differences between $u_{ri}^{s}$ and $v_{ri}^{s}$. For simplicity, we consider only a single group, $r$. Nodes that are colored in red are selected in $S_{n(r)}.$ Note that $u_{ri}^{s}$ and $v_{ri}^{s}$ are only defined for $\mathbf{y}_{ri} \in S_{n(r)}$.}
\label{fig:test}
\end{figure}

Let $\widetilde{D}_{ri}^s$ be the number of edges that $\mathbf{y}_{ri}$ has in $S_{n(s)}$ (excluding $\mathbf{y}_{ri}$ if $r = s$). For $\mathbf{y}_{ri}\in S_{n(r)}$, we define
\begin{align*}
u_{ri}^{s}  &  = I(D_{ri}^s \geq 1) \\
v_{ri}^{s}  &  = I(\widetilde{D}_{ri}^s \geq 1)
\end{align*}
so $u_{ri}^{s}$ is the indicator for an edge between $\mathbf{y}_{ri}\in S_{n(r)}$ with at least one node in $\Omega_s$ and $v_{ri}^{s}$ is a
\textquotedblleft sample version\textquotedblright\ of $u_{ri}^{s}$ with respect to $S_{n(s)}$. The differences between $u_{ri}^{s}$ and $v_{ri}^{s}$ is shown in Figure \ref{fig:test} . Note that $E(u_{ri}^{s}) = \theta_{rs}$. \\ 
\indent Nodes in $S_{n(r)}$, who do not link to any nodes in $S_{n(s)}$, may in fact be linked to nodes in $\Omega_s$ but were not observed in $S_{n(s)}$. Thus, $v_{ri}^{s} \leq u_{ri}^{s}$ for all $1\leq i\leq n_{r}$ and the estimator that ignores the impact of incomplete data, $\widetilde{\theta}_{rs} = \frac{1}{n_r} \sum_{i = 1}^{n_r}v_{ri}^s$, is biased downward (unless all of $\Omega$ is sampled):
\begin{equation}
E\left(  \widetilde{\theta}_{rs}\right)  =\frac{1}{n_{r}}\sum_{i=1}^{n_{r}%
}E\left( v_{i}^{rs}\right)  =E\left(v_{i}^{rs}\right)  \leq E\left(
u_{i}^{rs}\right)  =\theta_{rs}. 
\end{equation}

\noindent We refer to $\widetilde{\theta}_{rs}$ as the unadjusted estimator.

\section{Methods}

As shown in Section 3, the unadjusted estimator for the linkage rate is biased downwards for $\theta_{rs}$. Additionally, even if $\widetilde{\theta}_{rs}$ were an unbiased estimator,the Central Limit Theorem and Law of Large Numbers cannot be directly applied, because the independence assumption is violated. Hence, we use a U-Statistics framework to derive an estimator for $\theta_{rs}$ that is asymptotically normal and consistent for $\theta_{rs}$. 

\subsection{An Unbiased Estimator of Probability of Linkage}

First, we note that $v_{ri}^{s} = 1$ implies that $u_{ri}^{s} = 1$, because if a edge is observed in $S_n$, then it must exist in $\Omega$. It follows that 

\begin{align*}
E(\widetilde{\theta}_{rs}) &=  \Pr(v_{ri}^s = 1) \\
                    &=  \Pr(v_{ri}^s = 1, u_{ri}^s = 1) \\
                    &= \Pr(v_{ri}^s = 1 \mid u_{ri}^s = 1) \Pr(u_{ri}^s = 1) \\
                    &= \pi_{rs} \theta_{rs}
\end{align*}
where $\pi_{rs} = \Pr(v_{ri}^s = 1 \mid u_{ri}^s = 1)$, the probability of observing the edge between a node in $S_{n_{r}}$ with some node in $S_{n(s)}$ given that an edge does in fact exist between this node in $S_{n_{r}}$ and some node in $S_{n(s)}$. We have that 
\begin{align*}
\pi_{rs}  &=\Pr\left( v_{ri}^{s}=1\mid u_{ri}^{s}=1\right) \\
&  =\frac{\Pr\left( v_{ri}^{s}=1,u_{ri}^{s}=1\right)  }{\Pr\left(
u_{ri}^{s}=1\right)  }\\
&  = \frac{\Pr\left(v_{ri}^{s}=1 \right)  }{\Pr\left(  u_{ri}^{s}=1\right)
}%
\end{align*}
Therefore, the following is an unbiased estimator for $\theta_{rs}$:
\begin{equation}
\frac{1}{n_{r}\pi_{rs}}\sum_{i=1}^{n_{r}}v_{ri}^{s}, \label{eqn64}%
\end{equation}
However, in practice, $\pi_{rs}$  is unknown, because the event $\left\{  u_{i}%
^{rs}=1\right\}  $ is not observed. Thus, the above is not a feasible estimator for $\theta_{rs}$.

\subsection{A Feasible and Consistent Estimator for Linkage Rate}
Consider a  subsample from the nodes in $S_n$ of size $m=\sum_{r=1}^{w}p_{r}n_{r}$, which we denote by $S_m$, such that  for each group $r$ we randomly sample a proportion $p_r$ of the nodes in $S_{n(r)}$. We define $m_r = p_rn_r$. We denote the subsample from group $r$ as $S_{m(r)}.$ Note that $n_r = p_rN_r$ as well. Thus, in the subsample, we recapitulate the sampling of the observed data from the entire population. Let $\widetilde{\widetilde{D_{ri}^s}}$ be the number of nodes in $S_{m(s)}$ that have an edge with $\mathbf{y}_{ri}$ (excluding $\mathbf{y}_{ri}$ is $r = s$). For any $\mathbf{y}_{ri}\in S_{m(r)}$, we define
\begin{align*}
\widetilde{u}_{ri}^{s}  &  = I(\widetilde{D}_{ri}^s \geq 1) \\
\widetilde{v}_{ri}^{s}  &  = I(\widetilde{\widetilde{D_{ri}^s}} \geq 1)
\end{align*}

\indent We then denote $\widetilde{\pi}_{rs}$ as the following:

\begin{align*}
    \widetilde{\pi}_{rs} &= \Pr(\widetilde{v}_{ri}^{s}=1\mid \widetilde{u}_{ri}^{s} =1) \\ 
                         &= \frac{\Pr(\widetilde{v}_{ri}^{s}=1)}{\Pr(\widetilde{u}_{ri}^{s}=1)}
\end{align*}
\noindent We can then estimate $\widetilde{\pi}_{rs}$ by 

\begin{align*}
\widehat{\widetilde{\pi}}_{rs}  &  =\frac{\frac{1}{m_{r}}\sum_{i=1}^{m_{r}}\widetilde{v}_{ri}^{s}  }{\frac{1}{m_{r}}\sum_{i=1}^{m_{r}}\widetilde{u}_{ri}^{s}}
\end{align*}

\noindent which is well-defined based on $S_n$ as $\widetilde{v}_{ri}^{s}$ and $\widetilde{u}_{ri}^{s}$ are observed. \\ 
\indent We then want to show that as $N_{r}, N_{s} \rightarrow \infty$,
\begin{align*}
    \widetilde{\pi}_{rs} \rightarrow \pi_{rs}
\end{align*}
so that $\widehat{\widetilde{\pi}}_{rs}$ is a consistent estimator. 

\begin{theorem}
Suppose $\frac{\text{max}(D_{r1}^s, \ldots, D_{rN_r}^{s})}{\sqrt{N_{rs}}} \rightarrow 0$ as $N_{rs} \rightarrow \infty$. We note that $p_r$ and $p_s$ are fixed so $n_{r} \rightarrow \infty$ and $n_s \rightarrow \infty$ as $N_r \rightarrow \infty$ and $N_s \rightarrow \infty$, respectively. Suppose also that $\Pr(\widetilde{D}_{ri} = k \mid \widetilde{D}_{ri} \geq 1) \rightarrow \Pr(D_{ri}^s = k \mid D_{ri}^s \geq 1)$ as $N_{r}, N_s \rightarrow \infty$.
Then
\begin{align*}
\widetilde{\pi}_{rs}  &  =P(\widetilde{v}_{ri}^{s}=1\mid \widetilde{u}_{ri}^{s}=1)\rightarrow
P(v_{ri}^{s}=1\mid u_{ri}^{s}=1)=\pi_{rs},\quad\text{ as }N_{r}, N_s\rightarrow
\infty.
\end{align*}
\label{theorem1}
\end{theorem}

The proof of Theorem \ref{theorem1} is provided in the appendix. The first assumption is made in Section 2. Due to the assumption in Theorem \ref{theorem1} that $\Pr(\widetilde{D}_{ri} = k \mid \widetilde{D}_{ri} \geq 1) \rightarrow \Pr(D_{ri}^s = k \mid D_{ri}^s \geq 1)$ as $N_{r}, N_s \rightarrow \infty$, consistency requires further assumptions on the structure of our network. Since several investigators have notes that HIV genetic linkage networks appear to have this property and that a key goal with this paper is to make inference on the linkage rates for HIV genetic linkage networks, we assume that the degree distribution follows some power law distribution \cite{leigh2011, wertheim2017, wertheim2014}. For such distributions, there exists some $k_0$ such that for $k \geq k_0$ we have

\begin{align*}
    \Pr(D_{rs}^s = k \mid D_{rs}^s \geq 1) = \beta k^{-\alpha}
\end{align*}

\noindent where $2 \leq \alpha \leq 3$. Although Stumpf et. al. showed that this assumption will not hold theoretically with networks of power law distributions, we show in Section 4.4 that this assumption approximately holds for large enough values of $p_s$ resulting in consistent estimators. Theorem \ref{theorem1} shows that although $\{u_{ri}^{s}=1\}$ is not observed, we can develop a subsample $S_{m}$ of $S_{n}$ and estimate $\pi_{rs}$ by
treating $S_{n}$ as $\Omega$ and $S_{m}$ as $S_{n}$.\\
\indent Since $\widehat{\widetilde{\pi}}_{rs}$ applies only to a single subsample, the estimator can depend heavily on the specific subsample that was selected. Hence, we propose the following estimator for $\pi_{rs}:$
\[
\widehat{\pi}_{rs} = \frac{\binom{n_{r}}{m_{r}}^{-1}\binom{n_{s}}{m_{s}%
}^{-1}\sum_{S_{m(r)}\in C_{m(r)}^{n(r)}}\sum_{S_{m(s)}\in C_{m(s)}%
^{n(s)}}\frac{1}{m_{r}}\sum_{i=1}^{m_{r}} \widetilde{v}_{ri}^{s}
}{\binom{n_{r}}{m_{r}}^{-1}\binom{n_{s}}{m_{s}}^{-1}\sum_{S_{m(r)}\in
C_{m(r)}^{n(r)}}\sum_{S_{m(s)}\in C_{m(s)}^{n(s)}}\frac{1}{m_{r}}%
\sum_{i=1}^{m_{r}} \widetilde{u}_{ri}^{s} },
\]
where $C_{m(j)}^{n(j)}$ is the set of all possible combinations from
sampling $m_{j}$ nodes from $S_{n(j)}$. With such an estimator $\widehat{\pi}_{rs}$, we can consider a feasible estimator of  $\theta_{rs}$ as:
\begin{equation}
\widehat{\theta}_{rs}=\frac{1}{n_{r}\widehat{\pi}_{rs}}\sum_{i=1}^{n_{r}%
}v_{ri}^{s}, \label{eqn70}%
\end{equation}
 We refer to $\widehat{\theta}_{rs}$ as the adjusted estimator for $\theta_{rs}$. \\
\indent To establish consistency and asymptotic normality of the estimate in  (\ref{eqn70}), standard asymptotic methods such as the law of large numbers and central limit theorem cannot be directly applied. This is because $\widetilde{u}_{ri}^{s}, v_{ri}^{s}$ and  $\widetilde{v}_{ri}^{s}$ are not stochastically independent, thereby violating the required independence assumption. Below in Section 4.3, we describe an approach to establish such properties. 

\subsection{Inference on Linkage Rate: A U-Statistics Framework}
First, we let 
\begin{align*}
    \gamma_{rs1} &= \Pr\left(  \widetilde{v}_{ri}^{s}=1\right) \\ 
    \gamma_{rs2} &= \Pr\left(  \widetilde{u}_{ri}^{s}=1\right) \\ 
    \gamma_{rs3} &= \Pr\left(  v_{ri}^{s}=1\right). 
\end{align*}
 Now, we denote $\gamma_{rs}$ as the following:
\begin{align*}
    \gamma_{rs} = \left(
\begin{array}
[c]{c}%
\gamma_{rs1}\\
\gamma_{rs2}\\
\gamma_{rs3}%
\end{array}
\right).
\end{align*}
Then  $\frac{\gamma_{rs1}\gamma_{rs3}}{\gamma_{rs2}} \rightarrow \theta_{rs}$ as $N_{rs} \rightarrow \infty$ and by Theorem \ref{theorem1}, $\widetilde{\pi}_{rs} = \frac{\gamma_{rs1}}{\gamma_{rs2}}$.
From Section $4.1$ and $4.2$, we propose the following estimator for $\gamma$: 
\begin{align*}
      \widehat{\gamma}_{rs}\left(  \mathbf{y}_{r1},\mathbf{y}_{r2},\ldots,\mathbf{y}_{rn_{r}%
};\mathbf{y}_{s1},\mathbf{y}_{s2},\ldots,\mathbf{y}_{sn_{s}}\right)   &=  
\left(
\begin{array}
[c]{c}%
\widehat{\gamma}_1 \\ 
\widehat{\gamma}_2 \\ 
\widehat{\gamma}_3
\end{array}
\right) \\
 &= \left(
\begin{array}
[c]{c}%
\binom{n_{r}}{m_{r}}^{-1}\binom{n_{s}}{m_{s}%
}^{-1}\sum_{S_{m(r)}\in C_{m(r)}^{n(r)}}\sum_{S_{m(s)}\in C_{m(s)}%
^{n(s)}}\frac{1}{m_{r}}\sum_{i=1}^{m_{r}}\widetilde{v}_{ri}^{s}\\
\binom{n_{r}}{m_{r}}^{-1}\binom{n_{s}}{m_{s}}^{-1}\sum_{S_{m(r)}\in
C_{m(r)}^{n(r)}}\sum_{S_{m(s)}\in C_{m(s)}^{n(s)}}\frac{1}{m_{r}}%
\sum_{i=1}^{m_{r}}\widetilde{u}_{ri}^{s}\\
\frac{1}{n_{r}}\sum_{i=1}^{n_{r}%
}v_{ri}^{s} %
\end{array}
\right). 
\end{align*}
Thus, provided that we can establish consistency and asymptotic normality of $\widehat{\gamma}_{rs}$ , by the Delta Method and Theorem 1, $\widehat{\theta}_{rs}$ is  consistent and asymptotically normal. To this end, we adopt the following U-Statistics framework. \\ 
\indent First, we note that
\begin{align*}
    E(\widehat{\gamma}_{rs}) =  \gamma_{rs}
\end{align*}
and that the arguments of $\widehat{\gamma}_{rs}$ are invariant to permutations of nodes within each group, i.e., 
\begin{align*}
\widehat{\gamma}_{rs}\left(  \mathbf{y}_{r1^{\prime}},\mathbf{y}_{r2^{\prime}}%
,\ldots,\mathbf{y}_{rn_{r}^{\prime}};\mathbf{y}_{s1^{\prime \prime}},\mathbf{y}%
_{s2^{\prime \prime}},\ldots,\mathbf{y}_{sn_{s}^{\prime \prime}}\right)  = \widehat{\gamma}_{rs}\left(
\mathbf{y}_{r1},\mathbf{y}_{r2},\ldots,\mathbf{y}_{rn_{r}};\mathbf{y}%
_{s1},\mathbf{y}_{s2},\ldots,\mathbf{y}_{sn_{s}}\right) 
\end{align*}
where  $\left(  1^{\prime},2^{\prime},\ldots,n_{r}^{\prime}\right)$ and  $\left(  1^{\prime \prime},2^{\prime \prime},\ldots,n_{s}^{\prime \prime}\right)$ are any permutations of $(1, 2, \ldots, n_r)$ and $(1, 2, \ldots, n_s)$, respectively. Thus, by \cite{kowalski2007}, $\widehat{\gamma}_{rs}$ is a multivariate U-Statistic. 
Let
\begin{align*}
\mathbf{h}_{rs1}\left(  \mathbf{y}_{ki}\right)   &  =E\left(  \widehat{\gamma}_{rs}\left(
\mathbf{y}_{r1}\ldots,\mathbf{y}_{rn_r} ; \mathbf{y}_{s1}\ldots,\mathbf{y}_{sn_s} \right)  \mid\mathbf{y}_{ki}\right)
,\quad\widetilde{\mathbf{h}}_{rs1}\left(  \mathbf{y}_{ki}\right)  =\mathbf{h}%
_{rs1}\left(  \mathbf{y}_{ki}\right)  - \gamma_{rs},\\
\Sigma_{k}  &  =\text{Var}\left[  \widetilde{\mathbf{h}}_{rs1}\left(  \mathbf{y}%
_{ki}\right)  \right]  =E\left[  \widetilde{\mathbf{h}}_{rs1}\left(
\mathbf{y}_{ki}\right)  \widetilde{\mathbf{h}}_{rs1}^{\top}\left(  \mathbf{y}%
_{ki}\right)  \right]  .
\end{align*} 
By \cite{kowalski2007, sarstedt2018, tang2012}, it follows that 
\[
\sqrt{n_{rs}}\left(  \widehat{\gamma}_{rs}-\gamma_{rs}\right)  \rightarrow_{d}N\left(
\mathbf{0},\Sigma_{\gamma(rs)}\right)  .
\]
where 
\[ n_{rs}  = \begin{cases} 
      n_{r}  & r = s \\
      n_r + n_s & r \neq s \\
   \end{cases}.
\]
\begin{align*}
    \Sigma_{\gamma(rs)}=\frac{N_r - n_r}{N_s}\rho_r^2n_r^{2}\Sigma_{r} + \frac{N_s - n_s}{N_s} \rho_s^2n_s^{2}\Sigma_{s}
\end{align*}

\noindent and 
\begin{align*}
    \rho_{k}^{2}=\lim_{n_{rs}\rightarrow\infty}\frac{n_{rs}}{n_{k}}. 
\end{align*}
\noindent A consistent estimate of $\Sigma_{\gamma_{(rs)}}$ is given by:

\begin{align*}
\widehat{\Sigma}_{\gamma(rs)}&= \frac{N_r - n_r}{N_r} \frac{n_{rs}}{n_r}n_r^{2}\widehat{\Sigma}_{r} + \frac{N_s - n_s}{N_s}\frac{n_{rs}}{n_r} n_s^{2}\widehat{\Sigma}_{s} = n_{rs}(\frac{N_r - n_r}{N_r} n_r\widehat{\Sigma}_{r} + \frac{N_s - n_s}{N_s} n_s\widehat{\Sigma}_{s}), \\
\widehat{\Sigma}_{k} &= \frac{1}{n_k - 1} \sum_{i = 1}^{n_k} (\widehat{\mathbf{h}}_{rs1}(\mathbf{y}_{ki}) - \widehat{\gamma}_{rs})(\widehat{\mathbf{h}}_{rs1}(\mathbf{y}_{ki}) - \widehat{\gamma}_{rs})^T
\end{align*}
where $\widehat{\mathbf{h}}_{rs1}(\mathbf{y}_{ki})$ is a consistent estimate of $\mathbf{h}_{rs1}\left(  \mathbf{y}_{ki}\right)$. In the appendix, it is shown that $\widehat{\mathbf{h}}_{rs1}(\mathbf{y_{ri}})$ defined as follows is consistent:
\begin{align*}
\widehat{\mathbf{h}}_{rs1}(\mathbf{y}_{ri}) = \left(
\begin{array}
[c]{c}%
\frac{1}{n_r}  \binom{n_r-1}{m_r-1}^{-1}\sum_{S_{m(r)}\in C_{m(r)}^{n(r)}} \binom{n_s}{m_s}^{-1} \sum_{S_{m(s)}\in C_{m(s)}
^{n(s)}} \widetilde{v}_{ri}^{s} + \frac{n_r - 1}{n_r} \widehat{\gamma}_{rs1} \\
\frac{\widetilde{u}_{ri}^{s}}{n_r} + \frac{n_r - 1}{n_r} \widehat{\gamma}_{rs2} \\
\frac{\widetilde{u}_{ri}^{s}}{n_r} + \frac{n_r - 1}{n_r} \widehat{\gamma}_{rs3} 
\end{array}
\right)  
\end{align*}
\noindent and 
\begin{align*}
\widehat{\mathbf{h}}_{rs1}(\mathbf{y}_{si}) = \left(
\begin{array}
[c]{c}%
\widehat{\gamma}_{rs1} \\ 
\widehat{\gamma}_{rs2} \\ 
\widehat{\gamma}_{rs3} 
\end{array}
\right)  
\end{align*}
\noindent Thus, $\widehat{\Sigma}_s = 0$, which implies
\begin{align*}
    \widehat{\Sigma}_{\gamma(rs)} = n_{rs}n_r\widehat{\Sigma}_r.
\end{align*}

\begin{theorem}
Let
\begin{align*}
\mathbf{h}_{rs1}\left(  \mathbf{y}_{ki}\right)   &  =E\left(  \widehat{\gamma}_{rs}\left(
\mathbf{y}_{r1}\ldots,\mathbf{y}_{rn_r} ; \mathbf{y}_{s1}\ldots,\mathbf{y}_{sn_s} \right)  \mid\mathbf{y}_{ki}\right)  ,\quad \widetilde
{\mathbf{h}}_{rs}\left(  \mathbf{y}_{ki}\right)  =\mathbf{h}_{rs1}\left(
\mathbf{y}_{ki}\right)  -\gamma_{rs},\\
\Sigma_{k}  &  =Var\left[  \widetilde{\mathbf{h}}_{rs}\left(  \mathbf{y}%
_{ki}\right)  \right]  =E\left[  \widetilde{\mathbf{h}}_{rs}\left(
\mathbf{y}_{ki}\right)  \widetilde{\mathbf{h}}_{rs}^{\top}\left(
\mathbf{y}_{ki}\right)  \right]  ,\\
\quad n_{rs}  &  =%
\begin{cases}
n_{r} & \text{if }r=s\\
n_{r}+n_{s} & \text{if }r\neq s
\end{cases}
,\quad \rho_{k}^{2}=\lim_{n_{rs}\rightarrow \infty}\frac{n_{rs}}{n_{k}}%
<\infty,\quad k=r,s
\end{align*}
Then, we have:\ (1) $\widehat{\gamma}_{rs}$ is a consistent, unbiased and
asymptotically normal estimator of $\gamma_{rs}$:\
\[
\sqrt{n_{rs}}\left(  \widehat{\gamma}_{rs}-\gamma_{rs}\right)  \rightarrow
_{d}N\left(  \mathbf{0},\Sigma_{\gamma(rs)}=\frac{N_r - n_r}{N_r}\rho_{r}^{2}n_{r}^{2}\Sigma
_{r} + \frac{N_s - n_s}{N_s}\rho_{s}^{2}n_{s}^{2}\Sigma_{s}\right)  ,
\]
(2) A consistent estimator of the asymptotic variance is given by:
\begin{align*}
\widehat{\Sigma}_{\gamma(rs)}  &  =\frac{N_r - n_r}{N_r} \frac{n_{rs}}{n_{r}}n_{r}^{2}%
\widehat{\Sigma}_{r} + \frac{N_s - n_s}{N_s}\frac{n_{rs}}{n_{r}}n_{s}^{2}\widehat{\Sigma}_{s}%
=n_{rs}(\frac{N_r - n_r}{N_r} n_{r}\widehat{\Sigma}_{r} + \frac{N_s - n_s}{N_s}n_{s}\widehat{\Sigma}_{s}),\\
\widehat{\Sigma}_{k}  &  =\frac{1}{n_{k}-1}\sum_{i=1}^{n_{k}}\left(
\widehat{\mathbf{h}}_{rs1}\left(  \mathbf{y}_{ki}\right)  -\widehat{\gamma
}_{rs}\right)  \left(  \widehat{\mathbf{h}}_{rs1}\left(  \mathbf{y}%
_{ki}\right)  -\widehat{\gamma}_{rs}\right)  ^{\top},\\
\widehat{\mathbf{h}}_{rs1}\left(  \mathbf{y}_{ri}\right)   &  =\left(
\begin{array}
[c]{c}%
\frac{1}{n_{r}}\binom{n_{r}-1}{m_{r}-1}^{-1}\sum_{S_{m(r)}\in C_{m(r)}^{n(r)}%
}\binom{n_s}{m_s}^{-1} \sum_{S_{m(s)}\in C_{m(s)}^{n(s)}}\widetilde{v}_{ri}^{s}+\frac{n_{r}-1}{n_{r}}%
\widehat{\gamma}_{rs1}\\
\frac{\widetilde{u}_{ri}^{s}}{n_{r}}+\frac{n_{r}-1}{n_{r}}\widehat{\gamma}_{rs2}\\
\frac{\widetilde{u}_{ri}^{s}}{n_{r}}+\frac{n_{r}-1}{n_{r}}\widehat{\gamma}_{rs3}%
\end{array}
\right)  ,\quad \widehat{\mathbf{h}}_{rs1}(\mathbf{y}_{si})=\left(
\begin{array}
[c]{c}%
\widehat{\gamma}_{rs1}\\
\widehat{\gamma}_{rs2}\\
\widehat{\gamma}_{rs3}%
\end{array}
\right)  .
\end{align*}

\end{theorem}

With the asymptotic results in Theorem 2, we can readily obtain the
consistency and asymptotic normality by the Delta method. Let
\[
f\left(  \gamma_{rs}\right)  =\frac{\gamma_{rs1}\gamma_{rs3}}{\gamma_{rs2}}.
\]
Then, $\widehat{\theta}_{rs}=f\left(  \widehat{\gamma}_{rs}\right)  $. By the
Delta method and Theorem 2,
\[
\sqrt{n_{rs}}\left(  \widehat{\theta}_{rs}-\theta_{rs}\right)  \rightarrow
_{d}N\left(  \mathbf{0},\sigma_{\theta(rs)}^{2}=\phi^{\top}(\gamma_{rs}%
)\Sigma_{\gamma(rs)}\phi(\gamma_{rs})\right)  ,
\]
where
\[
\phi(\gamma_{rs})=\frac{\partial}{\partial \gamma_{rs}}f\left(  \gamma
_{rs}\right)  =\left(
\begin{array}
[c]{c}%
\frac{\partial f\left(  \gamma_{rs}\right)  }{\partial \gamma_{rs2}}\\
\frac{\partial f\left(  \gamma_{rs}\right)  }{\partial \gamma_{rs1}}\\
\frac{\partial f\left(  \gamma_{rs}\right)  }{\partial \gamma_{rs3}}%
\end{array}
\right)  =\left(
\begin{array}
[c]{c}%
-\frac{\gamma_{rs1}\gamma_{rs3}}{\gamma_{rs2}^{2}}\\
\frac{\gamma_{rs3}}{\gamma_{rs2}}\\
\frac{\gamma_{rs1}}{\gamma_{rs2}}%
\end{array}
\right)  .
\]
A consistent estimator of $\sigma_{\theta(rs)}^{2}$ is given by:\
\[
\widehat{\sigma}_{\theta(rs)}^{2}=\phi^{\top}\left(  \widehat{\gamma}%
_{rs}\right)  \widehat{\Sigma}_{\gamma_{rs}}\phi(\widehat{\gamma}_{rs}),
\]
where $\phi^{\top}\left(  \widehat{\gamma}_{rs}\right)  $ and $\widehat
{\Sigma}_{\gamma_{rs}}$ denote the respective quantities by substituting
$\widehat{\gamma}_{rs}$ in place of $\gamma_{rs}$. \ 

\subsection{Diagnostic}
The consistency of the proposed estimator depends on the assumption that $\Pr (\widetilde{D}_{ri}^s = k \mid \widetilde{D}_{ri}^s) \approx \Pr (D_{ri}^s = k \mid D_{ri}^s)$. Stumpf et. al. performed a simulation with sampling from scale-free networks \cite{stumpf2005}. They showed that the larger the value of $\alpha$, the greater the sample degree distribution deviates from the true distribution. Since for many settings $2 \leq \alpha \leq 3$ \cite{stumpf2005}, we apply our approach to a power law network with $\alpha = 3$ to evaluate values of $p$ such that 
\begin{align*}
\Pr (\widetilde{D}_{ri}^s = k \mid \widetilde{D}_{ri}^s) \approx \Pr (D_{ri}^s = k \mid D_{ri}^s)
\end{align*}

\begin{figure}
\includegraphics[scale=.55]{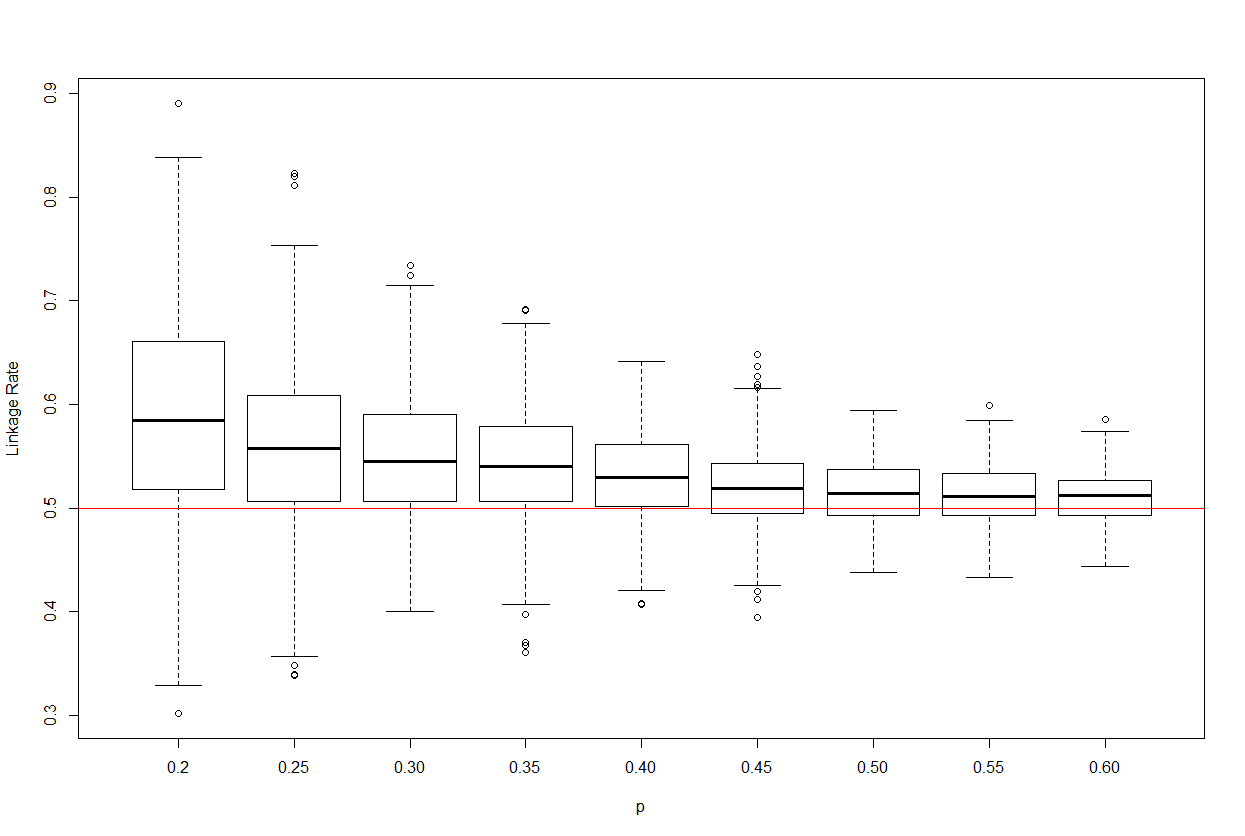}
\caption{Distribution of adjusted estimators for various values of $p$ when applying the proposed approach to a scale-free network with $\alpha = 3$. Here, the x-axis represents $p$ and the y-axis represents values for the adjusted estimators where the red line indicates the true value.}
\label{diagnosticplot}
\end{figure}

From Figure \ref{diagnosticplot}, we find that the estimates deviate greatly from the true value when $p < .40.$

\section{Applications}
\subsection{Simulation Study}
We apply the proposed methods to a population with two communities. We denote them both as community 1 and 2. We let $N_1 = 1000$ and $N_2 = 1200$. For the degree distributions, we have

    \[ \Pr (D_{ri}^r = k) = \begin{cases} 
          a k^{-2.5} & k \geq 1 \\
          0.50 & k = 0\\
       \end{cases}
    \]

    \[ \Pr (D_{ri}^s = k) = \begin{cases} 
          \beta k^{-2.6} & k \geq 1 \\
          0.60 & k = 0\\
       \end{cases}
    \]

    \[ \Pr (D_{si}^r = k) = \begin{cases} 
          \beta k^{-2.3} & k \geq 1 \\
          0.80 & k = 0\\
       \end{cases}
    \]

    \[ \Pr (D_{si}^s = k) = \begin{cases} 
          \beta k^{-3} & k \geq 1 \\
          0.70 & k = 0\\
       \end{cases}
    \]
\noindent We have that 
\[
\theta = 
  \begin{bmatrix}
    0.50 & 0.40  \\
    0.20 & 0.30 
  \end{bmatrix}
\]

\noindent We apply the proposed approach with letting $p_1 = 0.40$ and $p_2 = 0.60$. 

\begin{figure}
\includegraphics[scale=.47]{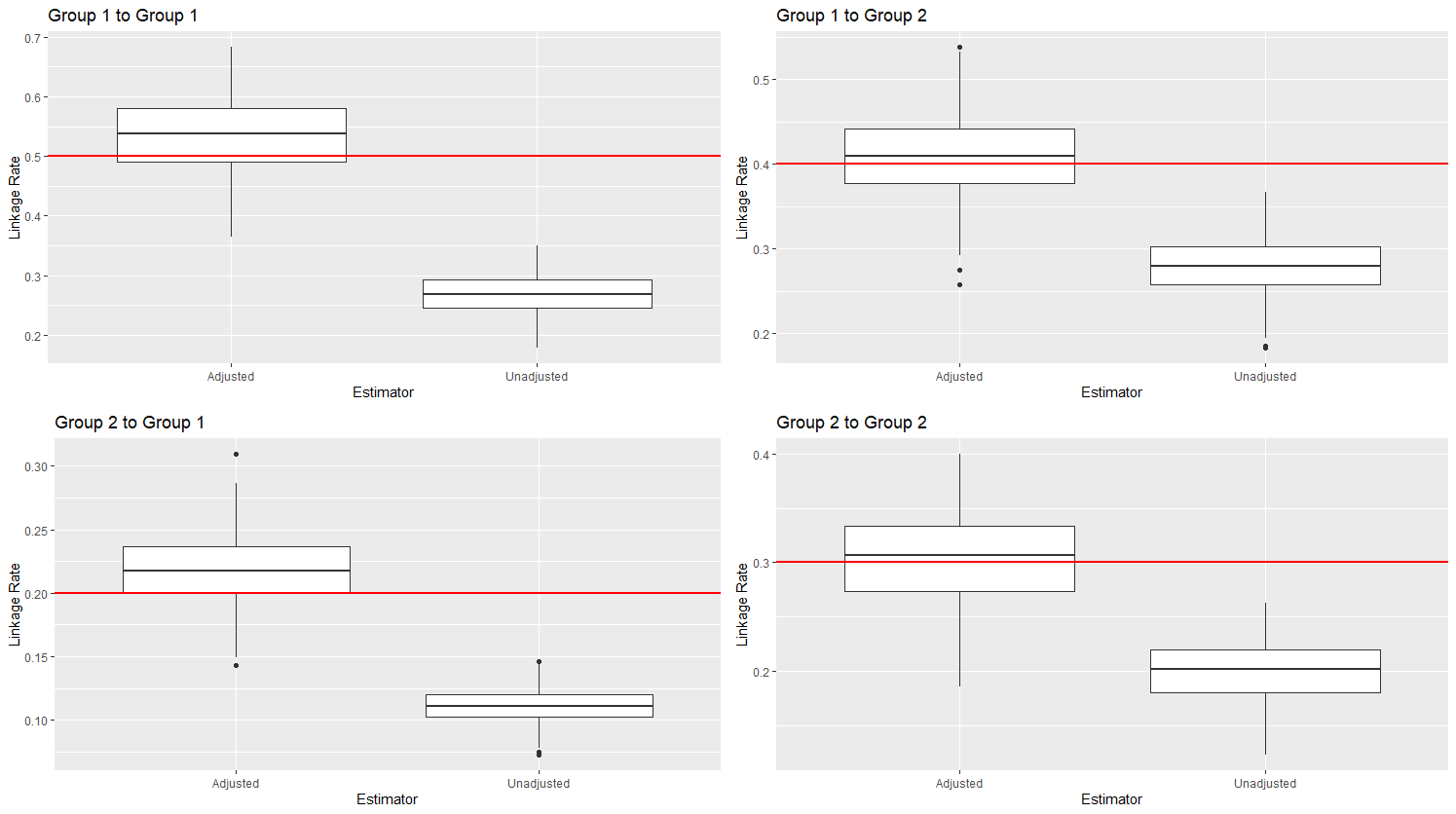}
\caption{Distribution of adjusted and unadjusted estimators for linkage rates from the simulation in Section 5.1. The red horizontal line is the true value. }
\label{powerlawcomm}
\end{figure}

 Figure \ref{powerlawcomm} demonstrates that the adjusted estimators are considerably less biases than are the unadjusted estimators. As expected, the sampling fraction of the group impacts the performance of the adjusted estimator. Figure \ref{powerlawcomm} confirms the expectation from Theorem 1 that when estimating the probability of linkage from Group "A" to Group "B", sampling additional nodes from Group "B" rather than Group "A" contributes more to improved performance. Most of the biases observed in  the adjusted estimators is upwards. \\ 
 \indent  Table \ref{coverageprob} shows that the coverage probabilities for the between community estimates are much greater than the estimates for within communities. Note that based on the definition of $\theta_{rs}$, it only makes sense to compare the coverage probability of $\theta_{11}$ with $\theta_{12}$. Similarly, the same case applies with $\theta_{21}$ and $\theta_{22}$.  
\begin{table}[H]
\centering
\begin{tabular}{|l|l|l|}
\hline
Group r & Group s & Coverage \\ \hline
1       & 1       & 0.68      \\ \hline
1       & 2       & 0.74       \\ \hline
2       & 1       & 0.64       \\ \hline
2       & 2       & 0.47       \\ \hline
\end{tabular}
\caption{Coverage probabilities of adjusted estimators from simulation in Section 5.1.}
\label{coverageprob}
\end{table}

\subsection{Botswana Combination Prevention Project}

As discussed earlier, the intent of developing this approach was to estimate linkage rates for the HIV viral genetic linkage networks Botswana. The data used comes from a large cluster-randomized trial of a combination HIV prevention intervention - the Botswana Combination Prevention Project (BCPP). \\ 
\indent In the BCPP, all households were targeted for a survey in 6 of the 30 participating communities in Botswana. The communities that were selected are Gumare, Mauntalala, Mmankgodi, Mmathethe, Ramokgonami and Shakawe. For those that choose to participate in the survey, demographic and household data along with HIV status is ascertained. For those who are HIV+, the viral genetic sequences are obtained. Hence, missing data arises due to the fact we have individuals who choose not to participate in the BCPP. However, going forward we assume that all individuals in the BCPP are MCAR, but acknowledge the fact that this assumption may not entirely hold.  
\\
\indent It follows that we define an edge between two individuals to exist if and only if the pairwise distance between their viral genetic sequences is below some threshold, $c$. Following Novitstky et. al, we set $c = 0.07$.
\\
\indent Table \ref{bcpp_prob} provides the proportions of HIV+ in individuals that participated in the BCPP; for 4 of the 6 communities, the proportions were over $40 \%$ but for 2, they were below $30 \%.$

\begin{table}[H]
\centering
\begin{tabular}{|l|l|l|l|l|l|l|}
\hline
  & Gumare & Maunatlala & Mmankgodi & Mmathethe & Ramokgonami & Shakawe \\ \hline
p & 0.29   & 0.52       & 0.26      & 0.44      & 0.48        & 0.48    \\ \hline
n & 325    & 363        & 270       & 336       & 350         & 484     \\ \hline
\end{tabular}
\caption{The proportion (p) and number (n) of HIV+ individuals in each community that participated in the BCPP. }
\label{bcpp_prob}
\end{table}

\begin{figure}
\includegraphics[scale=.4]{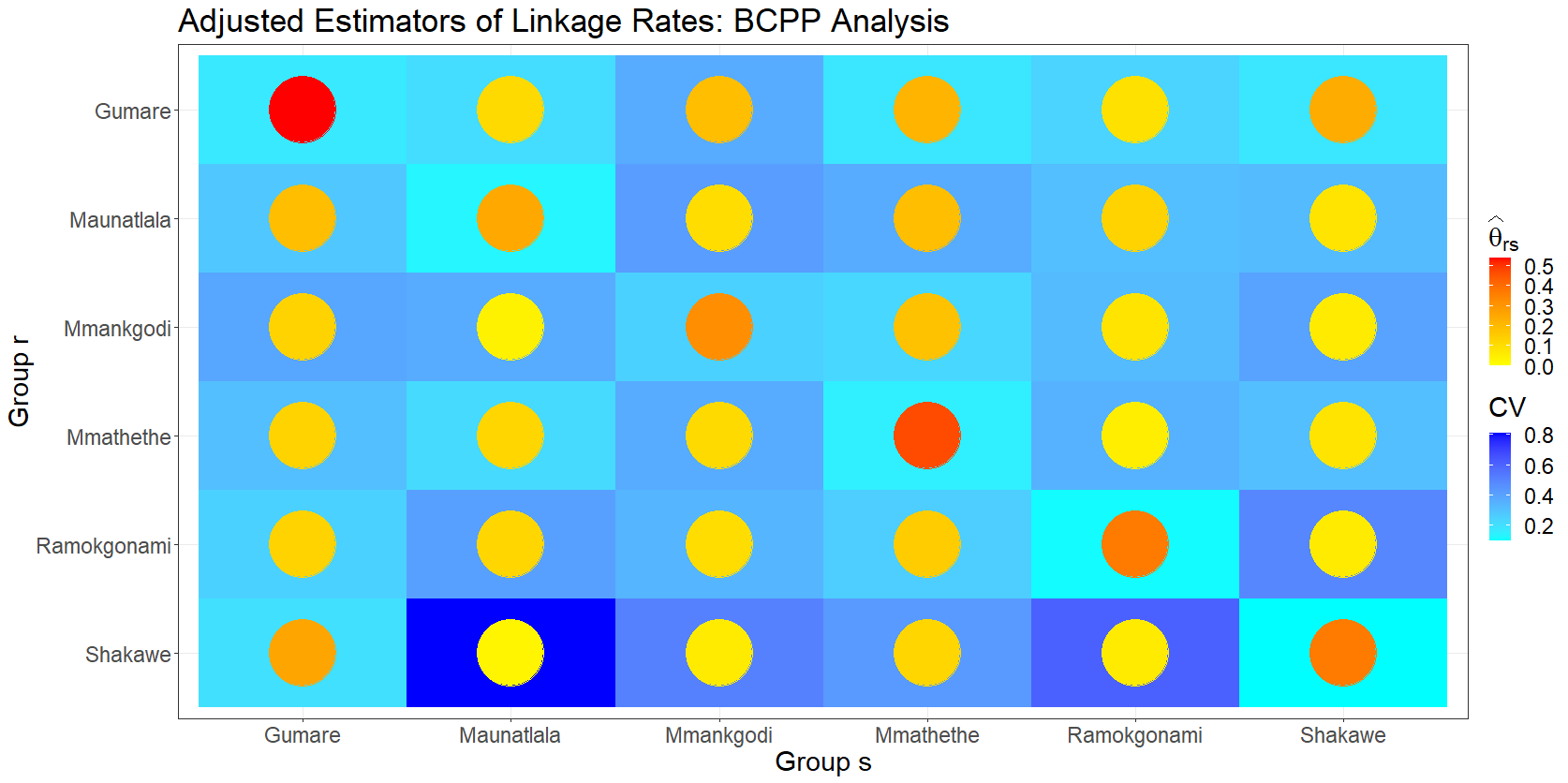}
\caption{Adjusted estimators for linkage rates between the communities from the BCPP and coefficients of variation (CV) indicated by colors of circle and cell, respectively. }
\label{bcpp_res}
\end{figure}

\indent We applied our methods to adjust for the incompleteness of the same. Figure \ref{bcpp_res} provides a heat map of the intensity of linkage after adjustment for missing data, within and across the communities as well as the variability associated with these estimates of linkage. For within community analyses, Gumare and Mmathethe have the highest linkage rates. Across communities, the high linkage rates are between Shakawe and Gumare in both directions. 

\section{Discussion}
This paper presents novel methods to estimate linkage rates in the presence of missing data. While methods have been proposed for such analyses, this paper is the first to ground such methods in statistical theory. Through the use of the U-statistics framework, we were able to show consistency of our estimator under assumptions about the nature of the network that are consistent with available literature and also to prove asymptotic normality, thereby permitting development of confidence interval estimates.  We demonstrate that the methods work well when sampling proportion is greater than 0.4; but even in a setting with lower sampling rates, the adjustment greatly improve performance of estimators compared to those that ignore missing data. How to make further improvements to our estimator when applied to data with low sampling rates is a topic for further research.

Our illustrative example made use of data from the HIV prevention study in Botswana—the BCPP.  We demonstrated that VGL linkage across communities is common—which implies that a treatment-as-prevention intervention applied at the village level will likely have effects on HIV incidence that are attenuated compared to effects that would occur if all relationships took place within villages.  Furthermore such estimates would also be attenuated compared to an estimand of interest—the counterfactual expected difference in incidence between a setting in which the intervention was implemented in all villages and a setting in which it in none.  Hence these VGL analyses are useful in both design and interpretation of cluster randomized trials for control of endemic diseases or disease outbreaks.

We note that our methods would apply not only to power law networks but to networks of all types, for which sampling of nodes is not complete.  Due to the interest in analyzing the BCPP, we chose our focus to be power law networks; future work is required for extension to networks more generally.

\section{Acknowledgements}
The BCPP Impact Evaluation was supported by the President’s Emergency Plan for AIDS Relief through the Centers for Disease Control and Prevention (CDC) (cooperative agreements U01 GH000447 and U2G GH001911). The contents of this article are solely the responsibility of the authors and do not necessarily represent the official positions of the funding agencies. The following grant supported Tyler Vu and Victor De Gruttola: NIAID  AI51164. We are grateful to Gabriel Erion for providing advice and code that greatly aided our work on the viral genetic sequences.
\newpage
\appendix
\section{Proof of Theorem \ref{theorem1}}

\begin{proof}
Without loss of generality suppose $p_sn_s \in \mathbb{N}$. If $p_sn_s \notin \mathbb{N}$, then we take $p_sn_s = \lceil p_sn_s \rceil$. We denote  $\widetilde{D}_{ri}^s$ to be the number of individuals in $S_{n(s)}$ linked to subject $\mathbf{y}_{ri} \in S_{n(r)}$, i.e., a sample version of \ $D_{ri}^s$. \ Note that \newline%
\[
P\left( \widetilde{D}_{ri}^s\geq1\mid D_{ri}^s\geq1 \right)  =P(v_{ri}^{s}=1\mid u_{ri}^{s}=1)=\pi_{rs}
\]
and 
\[
\widetilde{D}_{ri}^s\mid D_{ri}^s\geq1, D_{ri}^s = d \sim\text{HyperGeometric}\left(  N_s, d ,n_s\right)  
\]
where $N_s$ is the size of the population of group $s$, $d$ is the number of individuals in $\Omega_s$ that are linked to $\mathbf{y}_{ri} \in S_{n(r)}$ and $n_s$ is the size of the random sample taken from $\Omega_s$. \ We have%
\begin{align*}
P\left(  \widetilde{D}_{ri}^s \geq 1 \mid D_{ri}^s\geq1, D_{ri}^s = d \right)   &  =1-\frac
{\binom{N_s - d}{n_s}}{\binom{N_s}{n_s}}\\
&  =1-\frac{(N_s - d)!}{n_s!(N_s - d - n_s)!}\frac{n_s!(N_s - n_s)!}{N_s!}\\
&  =1-\frac{(N_s - d)!}{(N_s - d - n_s)!}\frac{(N_s - n_s)!}{N_s!} \\ 
&= 1 - \frac{\prod_{k = 0}^{n_s - 1}(N_s - k - d) }{\prod_{k = 0}^{n_s - 1}(N_s - k)} \\ 
&= 1 - \prod_{k = 0}^{n_s - 1} (1 - \frac{d}{N_s - k})
\end{align*}

\noindent  Take $A_N(d) = \prod_{k = 0}^{n_s - 1} (1 - \frac{d}{N_s - k})$, and let 
\begin{align*}
A_{N}^l(d) &= \log(A_N(d)) \\ 
&= \sum_{k = 0}^{n_s - 1} \log(1 - \frac{d}{N_s - k} )
\end{align*}
We note that when $x\rightarrow0,$ we have $\log(1+x)\rightarrow x+O(x^{2})$. For $N_s \rightarrow \infty$, We have 
\begin{align*}
A_N^l(d) &= - \sum_{k = 0}^{n_s}\frac{d}{N_s - k}\\ 
&=  - d \sum_{k = 0}^{n_s }\frac{1}{N_s - k}  \\ 
&= -d \sum_{j = (1 - p_s)N_s }^{N_s} \frac{1}{j}  \\ 
&=  -d \sum_{j = (1 - p_s)N_s }^{N_s}  \frac{1}{N_s} \frac{N_s}{j}\\ 
&= -d \int_{(1 - p_s)N_s }^{N_s} \frac{1}{N_s} \frac{N_s}{j} dj \\ 
&= -d \int_{1 - p_s}^{1} \frac{1}{x} dx \\ 
&= -d (- \log (1 - ps)) \\ 
&= d \log(1 - p_s).
\end{align*}

\noindent Thus, $A_N(d) \rightarrow (1 - p_s)^d$, and 

\begin{align*}
    P\left(  \widetilde{D}_{ri}^s \geq 1 \mid D_{ri}^s\geq1, D_{ri}^s = d \right) \rightarrow 1 - (1 - p_s)^d.
\end{align*}

\indent Let $D_{N(rs)}^{max} = \text{max}\{d: \Pr(D_{ri}^s = d ) \}$. It follows that we have
\begin{align*}
\pi_{rs} &= P\left(  \widetilde{D}_{ri}^s\geq1\mid D_{ri}^s\geq1 \right) \\
&= \sum_{d = 1}^{D_{N(rs)}^{max}}  P\left(  \widetilde{D}_{ri}^s\geq1\mid  D_{ri}^s = d \right) P(D_{ri}^s = d \mid D_{ri}^s\geq1) \\
&\rightarrow \sum_{d = 1}^{D^{max}_{N(rs)}} (1 - (1 - p_s)^d) P(D_{ri}^s = d \mid D_{ri}^s\geq1)
\end{align*}

\noindent Let $D_{n(rs)}^{max} = \text{max}\{d: \Pr(\widetilde{D}_{ri}^s = d) \}$. Similarly, if we treat $S_{n}$ as the population and $S_{m}$ as the
sample from $S_{n}$ of size $pn$, we can do the same as above and get the following:

\begin{align*}
    \widetilde{\pi}_{rs} &= \Pr(\widetilde{\widetilde{D^s_{rs}}} \geq 1 \mid \widetilde{D}_{ri}^s \geq 1) \\
    &= \sum_{d = 1}^{D^{max}_{n(rs)}} P(\widetilde{\widetilde{D^s_{rs}}} \geq 1 \mid \widetilde{D}_{ri}^s = d) P(\widetilde{D}_{ri}^s = d \mid \widetilde{D}_{ri}^s \geq 1) \\ 
    &\rightarrow \sum_{d = 1}^{D^{max}_{N(rs)}} (1 - (1 - p_s)^d) P(D_{ri}^s = d \mid D_{ri}^s \geq 1)
\end{align*}

Thus, $\widetilde{\pi}_{rs}$ is a consistent estimator for $\pi_{rs}$.

\end{proof}
\\
\section{Derivation of Consistent Estimators for $\mathbf{h}_{rs1}(\mathbf{y}_{ki})$} 
\subsection{A consistent estimator for $h_{rs1}^1(\mathbf{y}_{ki})$}

To find $\widehat{h}_{rs1}^{1}\left(  \mathbf{y}_{ki}\right)  $, a consistent estimator for  $h_{rs1}^1(\mathbf{y}_{ki})$, we first note
that
\begin{align*}
h_{rs1}^{1}\left(  \mathbf{y}_{ki}\right)   &  =E\left[  \widehat{\gamma}_{rs1}\left(
\mathbf{y}_{r1},\mathbf{y}_{r2},\ldots,\mathbf{y}_{rn_{r}};\mathbf{y}_{s1},\mathbf{y}_{s2},\ldots,\mathbf{y}_{sn_{s}}\right)  \mid\mathbf{y}_{ki}\right] \\
&  =\binom{n_{r}}{m_{r}}^{-1}\binom{n_{s}}{m_{s}}^{-1}\sum_{S_{m(r)}\in C_{m(r)}^{n(r)}}\sum_{S_{m(s)}\in C_{m(s)}%
^{n(s)}}\left\{  \frac{1}%
{m_{r}}\sum_{j=1}^{m_{r}}E\left[  \widetilde{v}_{ri}^s \mid\mathbf{y}_{ki}\right]  \right\} \\
&  =\binom{n_{r}}{m_{r}}^{-1}\binom{n_{s}}{m_{s}}^{-1}\sum_{S_{m(r)}\in C_{m(r)}^{n(r)}}\sum_{S_{m(s)}\in C_{m(s)}%
^{n(s)}}G_{ki}^{1}   ,
\end{align*}
where
\[
G_{ki}^1 = G_{ki}^{1}\left(  S_{m(r)}, S_{m(s)} \right)  =\frac{1}{m_{r}}\sum_{j=1}^{m_{r}}E\left[ \widetilde{v}_{ri}^s \mid\mathbf{y}_{ki}\right]  .\
\]
If $k = r$ and $\mathbf{y}_{ri} \in S_{m(r)}$, then
\[
G_{ki}^{1}=\frac{1}{m_{r}}\left[  \widetilde{v}_{ri}^s +\left(
m_r-1\right)  \gamma_{rs1}\right]  .
\]
Otherwise, 
\[
G_{ki}^{1}=\gamma_{rs1}.
\]

\noindent Therefore, we have $h_{rs1}^{1}\left(  \mathbf{y}_{si}\right) = \gamma_{rs1}$, which implies
\begin{align*}
\widehat{h}_{rs1}^{1}\left(  \mathbf{y}_{si}\right) = \widehat{\gamma}_{rs1}.    
\end{align*} 
For  $h_{rs1}^{1}\left(  \mathbf{y}_{ri}\right)$, we have there are $\binom{n_r-1}{m_r-1}$ subsets of size $m_r$ from $S_{n(r)}$ that contain $\mathbf{y}_{ri}$ and
the remaining $\binom{n_r-1}{m_r}$ subsets of $S_{n(r)}$ that do not contain $\mathbf{y}_{ri}$.
It follows that we derive the explicit form of $h_{rs1}^1\left(  \mathbf{y}_{ri}\right)$:
\begin{align*}
h_{rs1}^1\left(  \mathbf{y}_{ri}\right)   &  = \binom{n_{r}}{m_{r}}^{-1}\binom{n_{s}}{m_{s}}^{-1} \left[ \sum_{S_{m(r)}\in C_{m(r)}^{n(r)} : \mathbf{y}_{ri} \in S_{m(r)}} \sum_{S_{m(s)}\in C_{m(s)}^{n(s)}}  G_{ri}^1 +\sum_{S_{m(r)}\in C_{m(r)}^{n(r)} : \mathbf{y}_{ri} \notin S_{m(r)}} \sum_{S_{m(s)}\in C_{m(s)}^{n(s)}}  G_{ri}^1 \right] \\
&= \binom{n_{r}}{m_{r}}^{-1}\binom{n_{s}}{m_{s}}^{-1} \sum_{S_{m(r)}\in C_{m(r)}^{n(r)} : \mathbf{y}_{ri} \in S_{m(r)}} \sum_{S_{m(s)}\in C_{m(s)}^{n(s)}}  G_{ri}^1 +  \binom{n_{r}}{m_{r}}^{-1} \binom{n_r-1}{m_r} \gamma_1 \\ 
&= \binom{n_{r}}{m_{r}}^{-1}\binom{n_{s}}{m_{s}}^{-1} \sum_{S_{m(r)}\in C_{m(r)}^{n(r)} : \mathbf{y}_{ri} \in S_{m(r)}} \sum_{S_{m(s)}\in C_{m(s)}^{n(s)}}  \frac{1}{m_{r}}\left[   \widetilde{v}_{ri}^s +\left(
m_r-1\right)  \gamma_{1}\right]  +  \frac{n_r - m_r}{n_r} \gamma_1 \\ 
&= \binom{n_{r}}{m_{r}}^{-1}\binom{n_{s}}{m_{s}}^{-1}\sum_{S_{m(r)}\in C_{m(r)}^{n(r)} : \mathbf{y}_{ri} \in S_{m(r)}} \sum_{S_{m(s)}\in C_{m(s)}^{n(s)}}  \frac{1}{m_{r}}  \widetilde{v}_{ri}^s  + \binom{n_{r}}{m_{r}}^{-1}  \binom{n_r-1}{m_r-1} \frac{m_r - 1}{m_r} \gamma_1  +  \\&\frac{n_r - m_r}{n_r} \gamma_1  \\ 
&=  \binom{n_{r}}{m_{r}}^{-1}\binom{n_{s}}{m_{s}}^{-1} \sum_{S_{m(r)}\in C_{m(r)}^{n(r)} : \mathbf{y}_{ri} \in S_{m(r)}} \sum_{S_{m(s)}\in C_{m(s)}^{n(s)}}  \frac{1}{m_{r}}  \widetilde{v}_{ri}^s  + \frac{n_r - 1}{n_r} \gamma_1 \\ 
&= \left( \frac{1}{m_r}  \binom{n_{r}}{m_{r}}^{-1}\binom{n_r-1}{m_r-1} \right) \binom{n_r-1}{m_r-1}^{-1} \sum_{S_{m(r)}\in C_{m(r)}^{n(r)} : \mathbf{y}_{ri} \in S_{m(r)}} \binom{n_{s}}{m_{s}}^{-1}  \sum_{S_{m(s)}\in C_{m(s)}^{n(s)}}  \widetilde{v}_{ri}^s  + \frac{n_r - 1}{n_r} \gamma_1 \\
&= \frac{1}{n_r}  \binom{n_r-1}{m_r-1}^{-1} \sum_{S_{m(r)}\in C_{m(r)}^{n(r)} : \mathbf{y}_{ri} \in S_{m(r)}} \binom{n_{s}}{m_{s}}^{-1}  \sum_{S_{m(s)}\in C_{m(s)}^{n(s)}}  \widetilde{v}_{ri}^s  + \frac{n_r - 1}{n_r} \gamma_1
\end{align*}

\noindent Therefore,
\begin{align*}
\widehat{h}_{rs1}^1\left(  \mathbf{y}_{ri}\right)   = \frac{1}{n_r}  \binom{n_r-1}{m_r-1}^{-1} \sum_{S_{m(r)}\in C_{m(r)}^{n(r)} : \mathbf{y}_{ri} \in S_{m(r)}} \binom{n_{s}}{m_{s}}^{-1}  \sum_{S_{m(s)}\in C_{m(s)}^{n(s)}}  \widetilde{v}_{ri}^s  + \frac{n_r - 1}{n_r} \widehat{\gamma}_1
\end{align*}

\subsection{A consistent estimator for $h_{rs1}^2(\mathbf{y}_{ki})$}

\indent To find $\widehat{h}_{rs1}^{2}\left(  \mathbf{y}_{ki}\right)$, a consistent estimator for  $h_{rs1}^2(\mathbf{y}_{ki})$, we first note that
\begin{align*}
h_{rs1}^{2}\left(  \mathbf{y}_{ki}\right)   &  =E\left[  \widehat{\gamma}_{rs2}\left(
\mathbf{y}_{r1},\ldots,\mathbf{y}_{rn_{r}};\mathbf{y}_{s1},\ldots,\mathbf{y}_{sn_{s}}\right)  \mid\mathbf{y}_{ki}\right] \\
&  =\binom{n_{r}}{m_{r}}^{-1}\binom
{n_{s}}{m_{s}}^{-1}\sum_{S_{m(r)}\in C_{m(r)}^{n(r)}}\sum_{S_{m(s)}\in C_{m(s)}
^{n(s)}}\left\{ \frac{1}{m_{r}}\sum_{j=1}^{m_{r}}E[ \widetilde{u}_{ri}^s \mid \mathbf{y}_{ki}] \right\} \\
&  =\binom{n_{r}}{m_{r}}^{-1}\binom
{n_{s}}{m_{s}}^{-1}\sum_{S_{m(r)}\in C_{m(r)}^{n(r)}}\sum_{S_{m(s)}\in
C_{m(s)}^{n(s)}} G_{ki}^2 ,
\end{align*}
where
\[
G_{ki}^{2} = G_{ki}^{2}(S_{m(r)}, S_{n(s)}) = \frac{1}{m_{r}}\sum_{j=1}^{m_{r}}E[\widetilde{u}_{ri}^s \mid \mathbf{y}_{ki}] .\
\]

\noindent If $k = r$ and $\mathbf{y}_{ri} \in S_{m_r}$, then
\[
G_{ki}^{2} =\frac{1}{m_r}\left[  \widetilde{u}_{ri}^s +\left(  m_r-1\right)  \gamma_{rs2}\right]  .
\]
Otherwise,
\[
G_{ki}^{2} =\gamma_{rs2}.
\]

\noindent Therefore, we have $h_{rs1}^{2}\left(  \mathbf{y}_{si}\right) = \gamma_{rs2}$, which implies a consistent estimator $\widehat{h}_{rs1}^{2}\left(  \mathbf{y}_{si}\right)$ for $h_{rs1}^{2}\left(  \mathbf{y}_{si}\right)$ is defined as follows:
\begin{align*}
    \widehat{h}_{rs1}^{2}\left(  \mathbf{y}_{si}\right) = \widehat{\gamma}_{rs2}.
\end{align*}
Further, as $\widetilde{u}_{ri}^s$ is a connection indicator with
respect to $S_{n(s)}$,  as long as we know $\mathbf{y}_{ri} \in S_{m(r)}$ we have $\widetilde{u}_{ri}^s$ does not depend on $S_{m}$, i.e., if $S_{m(r)}$ and $S_{m(r)}^{\prime}$ are both subsamples of size $m_r$ from $S_{n(r)}$ that contains $\mathbf{y}_{ri}\in S_{m(r)} $, then
\begin{align*}
 G_{ki}^{2}(S_{m(r)}, S_{n(s)}) =G_{ki}^{2}(S_{m(r)}, S_{n(s)}).
\end{align*}
  For $h_{rs1}^{2}\left(  \mathbf{y}_{ri}\right)$,  we have $\binom{n_r-1}{m_r-1}$ subsets of size $m_r$ from $S_{n_r}$ that contain $\mathbf{y}_{ri}$
and there are the remaining $\binom{n_r-1}{m_r}$ subsets of size $m_r$ from $S_{n_r}$
that do not contain $\mathbf{y}_{ri}$. Fix $S_{m_r}^{\prime}$ to be any subset of size $m_r$
from $S_{n_r}$ that contain $\mathbf{y}_{ri}$. Thus,
\begin{align*}
h_{rs1}^2\left(  \mathbf{y}_{ri}\right)   &  = \binom{n_{r}}{m_{r}}^{-1}\binom{n_{s}}{m_{s}}^{-1} \left [ \sum_{S_{m(r)}\in C_{m(r)}^{n(r)} : \mathbf{y}_{ri} \in S_{m(r)}} \sum_{S_{m(s)}\in C_{m(s)}^{n(s)}}  G_{ri}^2 + \sum_{S_{m(r)}\in C_{m(r)}^{n(r)} : \mathbf{y}_{ri} \notin S_{m(r)}} \sum_{S_{m(s)}\in C_{m(s)}^{n(s)}}  G_{ri}^2 \right ] \\ 
&  = \binom{n_{r}}{m_{r}}^{-1} \binom{n_{r} - 1}{m_{r} - 1}  G_{ri}^2(S_{m_r^{\prime}}, S_{n(s)}) + \binom{n_{r}}{m_{r}}^{-1} \binom{n_r - 1}{m_r} \gamma_{rs2}  \\ 
&= \frac{m_r}{n_r}  G_{ri}^2(S_{m_r^{\prime}}, S_{n(s)}) + \frac{n_r - m_r}{n_r} \gamma_{rs2} \\
&= \frac{1}{n_r}\left[ \widetilde{u}_{ri}^s +\left(  m_r-1\right)  \gamma
_{rs2}\right] + \frac{n_r - m_r}{n_r} \gamma_{rs2}  \\ 
&= \frac{\widetilde{u}_{ri}^s}{n_r} + \frac{n_r - 1}{n_r} \gamma_{rs2}
\end{align*}
\noindent Therefore, 
\[
\widehat{h}_{rs1}^{2}\left(  \mathbf{y}_{ri}\right)  =\frac{\widetilde{u}_{ri}^s}{n_r} + \frac{n_r - 1}{n_r} \widehat{\gamma}_{rs2}
\]

\subsection{A consistent estimator for $h_{rs1}^3(\mathbf{y}_{ki})$}

To find $\widehat{h}_{rs1}^3(\mathbf{y}_{ki})$, a consistent estimator for $h_{1}^{3}(\mathbf{y}_{i})$, we first note that%

\begin{align*}
h_{1}^{3}(\mathbf{y}_{ki})  &  = E[\widehat{\gamma}_{rs3}(  \mathbf{y}_{r1}, \ldots,\mathbf{y}_{rn_{r}} ; \mathbf{y}_{s1},\ldots, \mathbf{y}_{sn_{s}}) \mid\mathbf{y}_{ki}]\\
&  = \frac{1}{n_r} \sum_{j = 1}^{n_r} E[v_{rj}^{s} = 1 \mid\mathbf{y}_{ki}]\\
&=   \left\{
    \begin{array}{c@{}c@{}c}
      \frac{1}{n_r} [v_{ni}^{rs} + (n_r - 1)\gamma_{rs3}], & k = r \\
       \gamma_{rs3} , & k = s
    \end{array}
  \right.
\end{align*}
If $k = r$ and $j = i$,  then 
\begin{align*}
    E[v_{rj}^{s} \mid \mathbf{y}_{ki}] = v_{ri}^s.
\end{align*}
Otherwise, 
\begin{align*}
    E[v_{rj}^{s}] = \gamma_{rs3}.
\end{align*}
Therefore, 
\begin{align*}
\widehat{h}_{1}^{3}(\mathbf{y}_{ki})   =\left\{
    \begin{array}{c@{}c@{}c}
      \frac{v_{ni}^{rs}}{n_r} + \frac{n_r - 1}{n_r}\gamma_{rs3}, & k = r \\
       \gamma_{rs3} , & k = s
    \end{array}
  \right.
\end{align*}
Thus, 
\begin{align*}
\widehat{h}_{1}^{3}(\mathbf{y}_{ki})   =\left\{
    \begin{array}{c@{}c@{}c}
     \frac{v_{ni}^{rs}}{n_r} + \frac{n_r - 1}{n_r}\widehat{\gamma}_{rs3}, & k = r \\
       \widehat{\gamma}_{rs3} , & k = s
    \end{array}
  \right.
\end{align*}

\section{A Consistent Estimator for the Variance of $\widetilde{\theta}_{rs}$}
Note that 
\begin{align*}
    \widetilde{\theta}_{rs} = \widetilde{\theta}_{rs}(  \mathbf{y}_{r1}, \ldots,\mathbf{y}_{rn_{r}} ; \mathbf{y}_{s1},\ldots, \mathbf{y}_{sn_{s}}) = \frac{1}{n_r}\sum_{i = 1}^{n_r}v_{ri}^s.  
\end{align*}
We have that the arguments of $\widetilde{\theta}_{rs}$ are symmetric when when permuted with respect to each group and that 
\begin{align*}
    E(\widetilde{\theta}_{rs}) = \pi_{rs}\theta.
\end{align*}
Thus, $\widetilde{\theta}_{rs}$ is a U-Statistic for $\pi_{rs}\theta$. Note that we denote $\widetilde{\theta}_{rs}$ and $\pi_{rs}\theta$ as $\widehat{\gamma}_3$ and $\gamma_3$, respectively. 
Let

\begin{align*}
    h_{rs1}^3 &=  E(h_{rs}^3\left(\mathbf{y}_{r1}, \ldots,\mathbf{y}_{rn_{r}} ;\mathbf{y}_{s1}, \ldots, \mathbf{y}_{sn_{s}}\right)  | \mathbf{y}_{ki}), \\ \widetilde{h}_{rs1}^{3}(\mathbf{y}_{i}) &= h_{rs1}^{3}(\mathbf{y}_{i}) - \gamma_{rs3} \\ 
    \sigma_{h(3)}^{2} &= Var(\widetilde{h}_{rs1}^{3}(\mathbf{y}_{ki}))
\end{align*}

\noindent By \cite{kowalski2007}, it follows that 

\[
\sqrt{n_{rs}}\left(  \widehat{\gamma}_{rs3}-\gamma_{rs3} \right)  \rightarrow
_{d}N\left(  0,\sigma_{\gamma(3)}^{2}= \rho_r^2 n_r^2 \sigma_{r3}^2 + \rho_s^2 n_s^2 \sigma_{s3}^2\right)  .
\]
where $\rho_{k}^{2}=\lim_{n_{rs}\rightarrow\infty}\frac{n_{rs}}{n_{k}}$ and 
\[ n_{rs}  = \begin{cases} 
      n_{r}  & r = s \\
      n_r + n_s & r \neq s \\
   \end{cases}.
\]

\noindent A consistent estimate of  $\sigma_{\gamma(3)}^{2}$ is given by:\
\begin{align}
\widehat{\sigma}_{\gamma(3)}^{2} &  =\frac{1}{n_{k}-1}\sum_{i=1}^{n_{k}}\left(  \widehat{h}_{rs1}^{3}\left(
\mathbf{y}_{ki}\right)  -\widehat{\gamma}_{rs3}\right)^{2},\nonumber
\end{align}
where $\widehat{h}_{rs1}^{3}\left(  \mathbf{y}_{ki}\right)  $ denotes a consistent
estimator for $h_{rs1}^{3}\left(  \mathbf{y}_{ki}\right)$.

\section{Comparison of Adjusted and Unadjusted Probabilities of Linkage }
\subsection{Estimates for Probabilities of Linkage}
\begin{table}[H]
\small
\begin{tabular}{|l|l|l|l|l|l|l|l|l|l|l|l|l|}
\hline
            & \multicolumn{2}{l|}{Gumare} & \multicolumn{2}{l|}{Maunatlala} & \multicolumn{2}{l|}{Mmankgodi} & \multicolumn{2}{l|}{Mmathethe} & \multicolumn{2}{l|}{Ramokgonami} & \multicolumn{2}{l|}{Shakawe} \\ \hline
            & U            & A            & U              & A              & U              & A             & U              & A             & U               & A              & U             & A            \\ \hline
Gumare      & 0.26         & 0.54         & 0.08           & 0.11           & 0.06           & 0.19          & 0.12           & 0.22          & 0.07            & 0.09           & 0.131         & 0.24         \\ \hline
Maunatlala  & 0.09         & 0.19         & 0.17           & 0.25           & 0.04           & 0.10          & 0.10           & 0.19          & 0.08            & 0.13           & 0.04          & 0.08         \\ \hline
Mmankgodi   & 0.06         & 0.13         & 0.03           & 0.04           & 0.10           & 0.32          & 0.09           & 0.18          & 0.04            & 0.08           & 0.03          & 0.06         \\ \hline
Mmathethe   & 0.05         & 0.13         & 0.07           & 0.12           & 0.04           & 0.11          & 0.26           & 0.47          & 0.03            & 0.05           & 0.05          & 0.08         \\ \hline
Ramokgonami & 0.06         & 0.13         & 0.07           & 0.12           & 0.03           & 0.10          & 0.08           & 0.15          & 0.23            & 0.37           & 0.03          & 0.06         \\ \hline
Shakawe     & 0.10         & 0.26         & 0.02           & 0.03           & 0.02           & 0.06          & 0.06           & 0.12          & 0.03            & 0.06           & 0.22          & 0.37         \\ \hline
\end{tabular}
\caption{\label{tab:table-name} Unadjusted (U) and adjusted (A) linkage rates for the communities investigated from the BCPP. }
\end{table}

\subsection{Standard Error of Probabilities of Linkage}

\begin{table}[h]
\small
\begin{tabular}{|l|l|l|l|l|l|l|l|l|l|l|l|l|}
\hline
            & \multicolumn{2}{l|}{Gumare} & \multicolumn{2}{l|}{Maunatlala} & \multicolumn{2}{l|}{Mmankgodi} & \multicolumn{2}{l|}{Mmathethe} & \multicolumn{2}{l|}{Ramokgonami} & \multicolumn{2}{l|}{Shakawe} \\ \hline
            & U            & A            & U              & A              & U              & A             & U              & A             & U               & A              & U             & A            \\ \hline
Gumare      & 0.13         & 0.09         & 0.07           & 0.02           & 0.02           & 0.07          & 0.02           & 0.04          & 0.02            & 0.02           & 0.20          & 0.04         \\ \hline
Maunatlala  & 0.02         & 0.05         & 0.03           & 0.03           & 0.03           & 0.04          & 0.08           & 0.07          & 0.01            & 0.04           & 0.02          & 0.03         \\ \hline
Mmankgodi   & 0.17         & 0.05         & 0.01           & 0.02           & 0.03           & 0.08          & 0.03           & 0.04          & 0.02            & 0.03           & 0.01          & 0.02         \\ \hline
Mmathethe   & 0.16         & 0.04         & 0.04           & 0.03           & 0.01           & 0.04          & 0.09           & 0.07          & 0.04            & 0.02           & 0.06          & 0.02         \\ \hline
Ramokgonami & 0.02         & 0.03         & 0.04           & 0.05           & 0.03           & 0.03          & 0.08           & 0.04          & 0.09            & 0.04           & 0.04          & 0.03         \\ \hline
Shakawe     & 0.03         & 0.05         & 0.02           & 0.03           & 0.02           & 0.03          & 0.07           & 0.05          & 0.04            & 0.03           & 0.34          & 0.04         \\ \hline
\end{tabular}
\caption{\label{tab:table-name} Standard errors of unadjusted (U) and adjusted (A) linkage rates for the communities investigated from the BCPP.}
\end{table}

\end{document}